\documentclass[twoside]{GleghornLab-StyleBioRxiv}
\usepackage{blindtext}
\usepackage{hyperref}
\usepackage{amsmath}
\usepackage{amssymb}
\usepackage[many]{tcolorbox}
\usepackage{listings}
\usepackage{subcaption}
\usepackage{float}
\usepackage{calc}
\usepackage{multirow}
\usepackage{booktabs}
\usepackage[table,dvipsnames]{xcolor}
\usepackage{algorithm}
\usepackage{algpseudocode}
\usepackage{tikz}
\usepackage{comment}
\usetikzlibrary{arrows.meta, positioning, calc, decorations.pathreplacing}

\leadauthor{Hallee}

\begin{document}

\title{Dual Triangle Attention: Effective Bidirectional Attention Without Positional Embeddings}
\shorttitle{Dual Triangle Attention}

\author[1, 2]{Logan Hallee}
\author[1, 2, 3 \Letter]{Jason P. Gleghorn}
\affil[1]{{Center for Bioinformatics and Computational Biology, University of Delaware}}
\affil[2]{{Synthyra, Newark, DE}}
\affil[3]{{Department of Biomedical Engineering, University of Delaware}}

\maketitle


\begin{abstract}
Bidirectional transformers are the foundation of many sequence modeling tasks across natural, biological, and chemical language domains, but they are permutation-invariant without explicit positional embeddings. In contrast, unidirectional attention inherently encodes positional information through its triangular mask, enabling models to operate without positional embeddings altogether. Here, we introduce \textbf{Dual Triangle Attention}, a novel bidirectional attention mechanism that separates the query-key subspace of each attention head into two complementary triangular masks: one that attends to past-and-self positions and one that attends to future-and-self positions. This design provides bidirectional context while maintaining the causal mask's implicit positional inductive bias in both directions. Using PyTorch's \texttt{flex\_attention}, Dual Triangle Attention is implemented as a single compiled kernel call with no additional parameters beyond standard multi-head attention. We evaluated Dual Triangle Attention across three settings: (1) a synthetic argmax position probe, (2) masked language modeling (MLM) on natural language, and (3) MLM on protein sequences. In the argmax task, both Dual Triangle Attention and causal attention learn positional information without explicit positional embeddings, whereas standard bidirectional attention cannot. In the MLM experiments, Dual Triangle Attention with Rotary Positional Embeddings (RoPE) achieved the best context extension performance and strong performance across the board. These findings suggest that Dual Triangle Attention is a viable attention mechanism for bidirectional transformers, with or without positional embeddings.
\end{abstract}



\section*{Introduction}

Bidirectional transformers are the foundation of representation learning across natural language processing, protein science, and beyond \cite{vaswani_attention_2017, brown_language_2020, lin_evolutionary-scale_2023, jumper_highly_2021}. In search, retrieval, and classification, each token benefits from attending to all other tokens, producing rich contextual embeddings \cite{warner_smarter_2024, hallee_contrastive_2025, hallee_diffusion_2025}. This full-context access is especially important for protein sequence modeling. In proteins, the 1D primary sequence is an abstract projection of a 3D molecule, so residue $i$ is frequently related to residue $j$ ($j > i$) through long-range spatial interactions, and vice versa \cite{lin_evolutionary-scale_2023, jumper_highly_2021}. Bidirectional protein language models have been applied to tasks including protein structure prediction \cite{lin_evolutionary-scale_2023, jumper_highly_2021, abramson_accurate_2024},  protein-protein interaction prediction \cite{hallee_protein-protein_2023, hallee_protein_2025, ko_tuna_2024}, direct functional annotation \cite{hallee_annotation_2024, su_saprot_2023}, codon-aware sequence modeling \cite{hallee_cdsbert_2023, outeiral_codon_2024}, and protein design/generation \cite{hallee_diffusion_2025, wang_diffusion_2024}, further underscoring the centrality of bidirectional attention in biological domains. Similar considerations apply to nucleotide modeling, where functional and structural dependencies can span large genomic distances \cite{nguyen_sequence_2024, brixi_genome_2025, avsec_advancing_2026}. Unfortunately, bidirectional attention has a fundamental vulnerability: because every token attends to every other token symmetrically, the mechanism is invariant to permutations of the input tokens \cite{vaswani_attention_2017, dufter_position_2021}. Without an explicit mechanism to encode token ordering, a bidirectional transformer cannot distinguish ``the dog chases the cat'' from ``the cat chases the dog.'' 

Resolving this permutation invariance has driven a progression of positional encoding strategies. Vaswani et al. introduced fixed sinusoidal encodings \cite{vaswani_attention_2017}, which were replaced by learned absolute embeddings in BERT \cite{devlin_bert_2019}. Shaw et al. proposed relative position representations that encode pairwise distances rather than absolute indices \cite{shaw_self-attention_2018}, a principle extended by T5's relative position biases \cite{raffel_exploring_2023} and by DeBERTa's disentangled position-content attention terms \cite{he_deberta_2021, he_debertav3_2023}. Rotary Positional Embeddings (RoPE) encode relative positions by rotating query and key vectors in the complex plane \cite{su_roformer_2023} and have become the dominant approach for both causal and bidirectional architectures \cite{touvron_llama_2023, grattafiori_llama_2024, jiang_mistral_2023, warner_smarter_2024, lin_evolutionary-scale_2023}. Recent alternatives include xPos, designed for length extrapolation \cite{sun_length-extrapolatable_2022}, and CoPE (Contextual Position Encoding), which learns position representations conditioned on content \cite{golovneva_contextual_2024}. However, when inference sequences exceed the training context length, RoPE's rotational phases become out-of-distribution, causing sharp performance degradation. Position Interpolation \cite{chen_extending_2023}, YaRN \cite{peng_yarn_2023}, and LongRoPE \cite{ding_longrope_2024} rescale rotational frequencies to accommodate longer sequences, but all require either expensive long-context finetuning or accept degraded performance at certain distances \cite{barbero_round_2024}.

A parallel line of research has questioned whether positional embeddings are strictly necessary for all transformer architectures. Haviv et al. first demonstrated that causal transformers without positional encoding can still learn positional information, with the triangular attention mask itself serving as an implicit positional signal \cite{haviv_transformer_2022}. Wang et al. extended this observation, showing that NoPE (\textbf{N}o \textbf{P}ositional \textbf{E}ncoding) transformers with causal masking can generalize to sequence lengths beyond training \cite{wang_length_2024}. Kazemnejad et al. formalized the mechanism, demonstrating that the first attention layer of a NoPE transformer can perfectly reconstruct absolute positions from the causal mask's triangular structure, with subsequent layers able to emulate any positional scheme \cite{kazemnejad_impact_2023}. Yang et al. proposed RoPE-to-NoPE, interleaving RoPE and NoPE layers in causal models to improve length generalization \cite{yang_rope_2025}. These results support the theoretical understanding of why causal NoPE models succeed and why bidirectional NoPE models cannot. 

The practical implications of this asymmetry became evident with DroPE (\textbf{P}ositional \textbf{E}mbeddings), which demonstrated that RoPE can be removed from pretrained autoregressive language models after a brief recalibration phase, yielding zero-shot context extension superior to RoPE-scaling methods \cite{gelberg_extending_2025}. The potential with DroPE-like methods is immense, since inference over very large contexts typically requires the much more expensive training over large contexts. DroPE's success rests on the assumptions that rotary embeddings are initially useful and that the causal mask can be exploited to recover performance after PE removal. However, this technique does not directly apply to bidirectional (BERT-like) transformers, since dropping position encoding will completely remove the models positional information. While proteins are typically shorter than modern natural-language contexts, modeling large pathways, whole proteomes, or entire genomes can still require very long context windows that DroPE may enable. For example, AlphaGenome and related models addressed megabase-scale DNA modeling with iterative convolutions that expand the receptive field after encoding instead of modeling the entire object in context due to computational constraints \cite{avsec_advancing_2026, linder_predicting_2025, boshar_foundational_2025}. Long-context requirements are equally relevant in NLP: modern LLM workflows often involve large corpora or conversation histories with documents that can span millions of tokens, where high-quality bidirectional embeddings are valuable for retrieval-augmented systems.

In this work, we propose \textbf{Dual Triangle Attention} (DTA), a novel bidirectional attention mechanism that introduces native position awareness into bidirectional attention. DTA splits the query-key subspace of each attention head into two complementary halves, each governed by a triangular mask. One half attends to the lower triangle (past positions and self), mirroring the structure of causal attention. The other half attends to the upper triangle (future positions and self), providing the complementary forward-looking context, with the diagonal included in both triangles. Critically, DTA introduces no additional parameters beyond standard multi-head attention and is implemented as a single compiled \texttt{flex\_attention} call using PyTorch's block mask API \cite{dong_flex_2024, paszke_pytorch_2019}, making it hardware-efficient and straightforward to integrate into existing architectures.

We evaluated DTA across three experimental settings, with the long-term goal of enabling robust context extension in bidirectional transformers:
\begin{enumerate}
    \item a synthetic argmax position probe that directly tests whether an attention mechanism can learn positional information
    \item masked language modeling on natural language text from FineWeb-Edu \cite{penedo_fineweb_2024}
    \item masked language modeling on protein sequences from OMG-Prot50 \cite{cornman_omg_2024}
\end{enumerate}
In each setting, we compared DTA against standard bidirectional and causal attention across multiple positional encoding strategies, including NoPE, learned absolute embeddings, RoPE, and DroPE-style position dropping. We hypothesized that complementary triangular masks, splitting each attention head's query-key subspace into forward-looking and backward-looking halves, provide sufficient directional structure for implicit positional encoding in bidirectional transformers; thereby enabling position awareness without PE, modeling capabilities on bidirectionally-native tasks like masked language modeling (MLM), and improved context extension with DroPE and/or RoPE. Across these experiments, DTA encodes positional information without explicit PE, matches or exceeds vanilla bidirectional performance on MLM, and offers favorable efficiency with an efficient compiled kernel, using roughly half the bidirectional attention FLOPs in the attention operation.


\section*{Results}

\begin{figure*}[b]
  \centering
  \includegraphics[width=.8\textwidth]{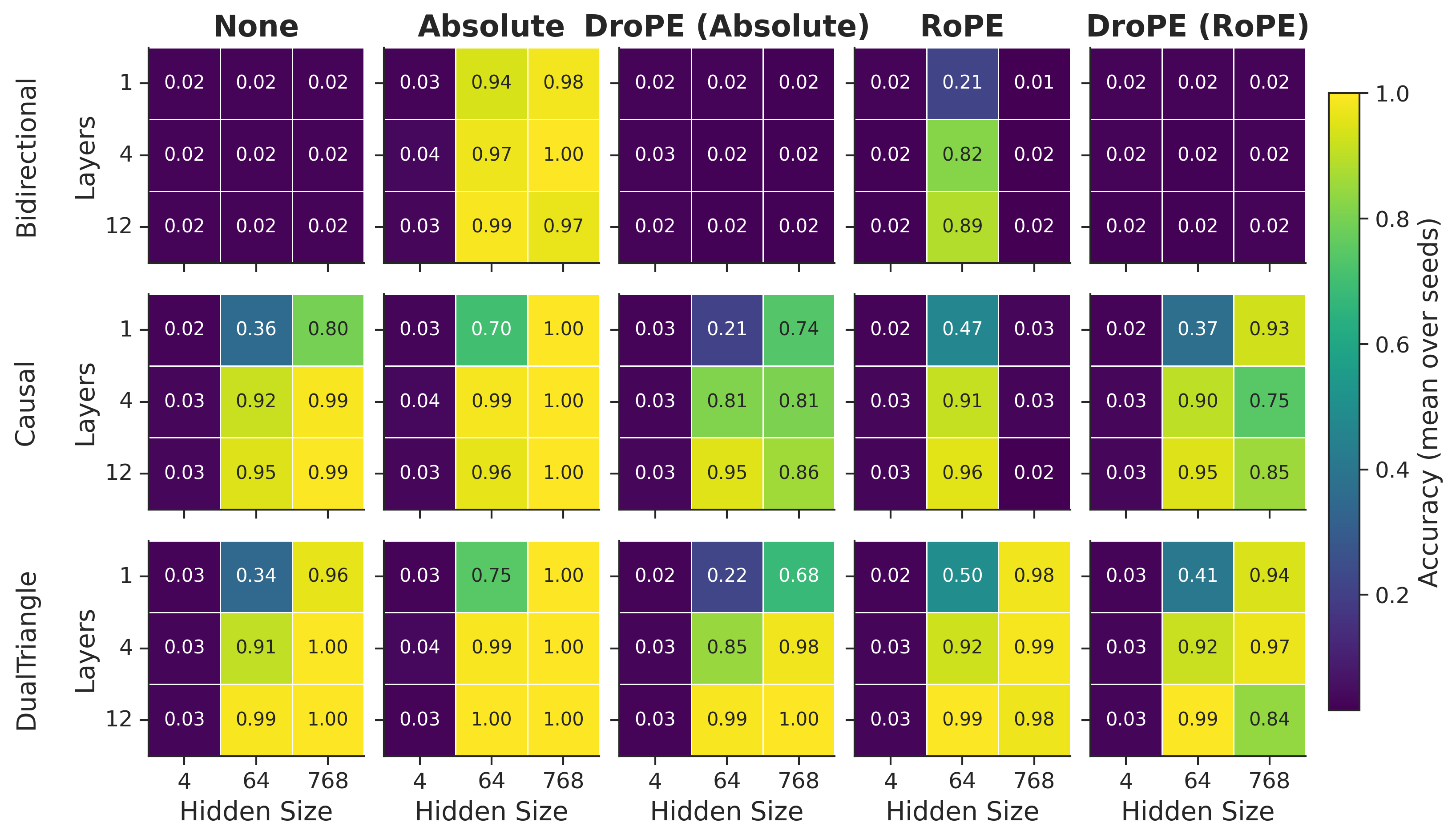}
  \caption{Argmax position probe accuracy across attention types, positional encoding strategies, and model configurations. Each cell shows the best evaluation accuracy (mean across three seeds) for a given combination of hidden size (columns) and number of layers (rows). Dual Triangle Attention and causal attention achieve high accuracy across nearly all configurations, including without explicit positional embeddings. Bidirectional attention fails completely without explicit positional embeddings.}
  \label{fig:argmax_heatmap}
\end{figure*}

\begin{figure*}[t]
  \centering
  \includegraphics[width=.75\textwidth]{figures/argmax/argmax_combined.png}
  \caption{Argmax probe accuracy trend, varying number of layers and hidden size. (a) Without positional embeddings, Dual Triangle and causal attention learn the task, while bidirectional attention cannot. Error bars show 95\% confidence intervals across three seeds. (b) At the largest model size (768 hidden, 12 layers), Dual Triangle Attention matches or exceeds both causal and bidirectional attention across all positional encoding strategies.}
  \label{fig:argmax_key}
\end{figure*}

\subsection*{Dual Triangle Attention encodes positional information without explicit embeddings}

To directly test whether different attention mechanisms can learn positional information, we designed a synthetic argmax position probe task. Given a sequence of random token IDs drawn uniformly from a fixed vocabulary, the model must predict the 0-indexed position of the maximum-valued token. This task requires the model to both compare token magnitudes and associate each token with its position in the sequence, making it a clean test of positional awareness.

We trained small transformer models across a grid of hidden sizes (4, 64, 768) and depths (1, 4, 12), comparing bidirectional, causal, and DTA under five positional encoding conditions: no positional embeddings, learned absolute, learned absolute with mid-training dropout (DroPE-style), RoPE, and RoPE with mid-training dropout (DroPE). Each configuration was trained with three seeds and evaluated on withheld data (\textbf{Figure \ref{fig:argmax_heatmap}}).

Across all positional encoding conditions where explicit positional information was available (learned absolute and RoPE), all three attention types achieved high accuracy, confirming that the task is well-posed and solvable by the transformer architecture. The critical distinction emerged in the \textit{no positional embeddings} condition. Causal attention maintained strong performance without any positional embeddings, consistent with prior findings that the causal mask implicitly encodes position \cite{haviv_transformer_2022, kazemnejad_impact_2023, wang_length_2024}. DTA similarly maintained strong performance, demonstrating that its complementary triangular masks provide sufficient positional signal in the bidirectional setting. In contrast, bidirectional attention failed entirely without positional embeddings, achieving accuracy no better than random chance.

\textbf{Figure \ref{fig:argmax_key}a} isolates the no-positional-embeddings condition across model sizes, clearly showing the dichotomy between attention types with triangular structure (causal, DTA) and those without (bidirectional). \textbf{Figure \ref{fig:argmax_key}b} compares all three attention types at the largest model size (768 hidden, 12 layers) across every positional condition, demonstrating that DTA is competitive with or superior to both baselines regardless of the positional encoding strategy. Interestingly, the causal + RoPE variant at the largest model size did not model the argmax task well, whereas the smaller variants did.

The DroPE-style conditions (learned absolute with mid-training dropout and RoPE with mid-training dropout) yielded results similar to their non-dropped counterparts for causal and DTA, as the triangular mask structure provides a fallback positional signal. For bidirectional attention, dropping positions mid-training predictably degraded performance because no implicit positional signal remains to compensate.

\subsection*{Masked language modeling on natural language}

Having established that DTA enables transformers to learn positional information, we next evaluated its performance on a practical task: masked language modeling (MLM) on natural language text. We trained 12-layer, 768-hidden size U-Net-style transformers \cite{ronneberger_u-net_2015, bao_all_2023, jordan_kellerjordanmodded-nanogpt_2026, hallee_logan_gleghorn-labspeedrunningplms_2026} from scratch on FineWeb-Edu \cite{penedo_fineweb_2024} using a custom BPE tokenizer with a vocabulary of 4,096 tokens \cite{sennrich_neural_2016}. Models were trained on 1 billion non-padding tokens with a training sequence length of 256 and evaluated at both the training context length (256 tokens, ``short'') and an extended context length (1,024 tokens, ``long'') to assess context extension capabilities. We compared all three attention types across three positional conditions: RoPE, no positional embeddings, and RoPE-off (DroPE-style: train with RoPE for 70\% of tokens, then remove). Each condition was run with three seeds.

\begin{figure*}[b]
  \centering
  \includegraphics[width=.78\textwidth]{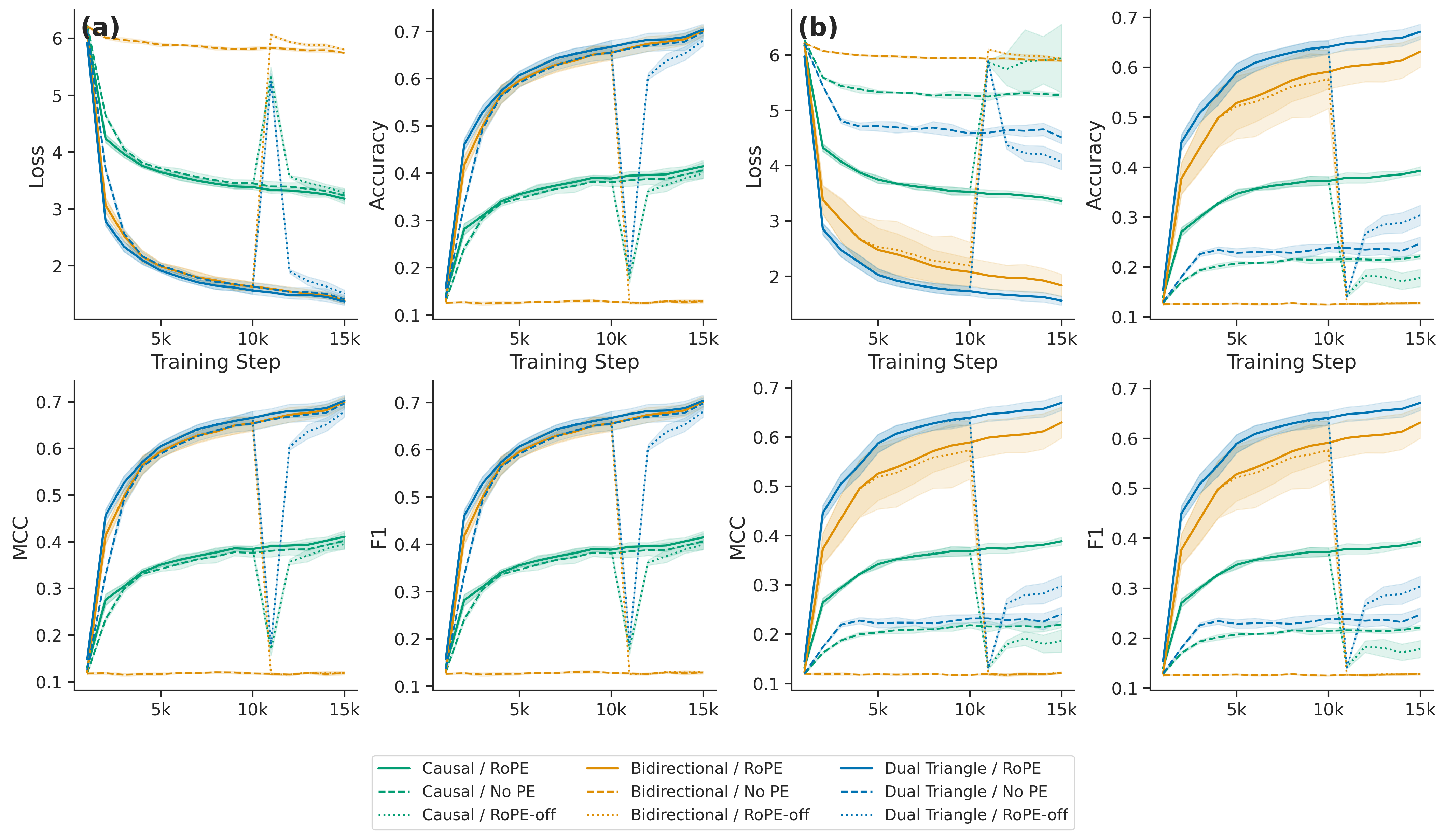}
  \vspace{0.5em}
  \includegraphics[width=.78\textwidth]{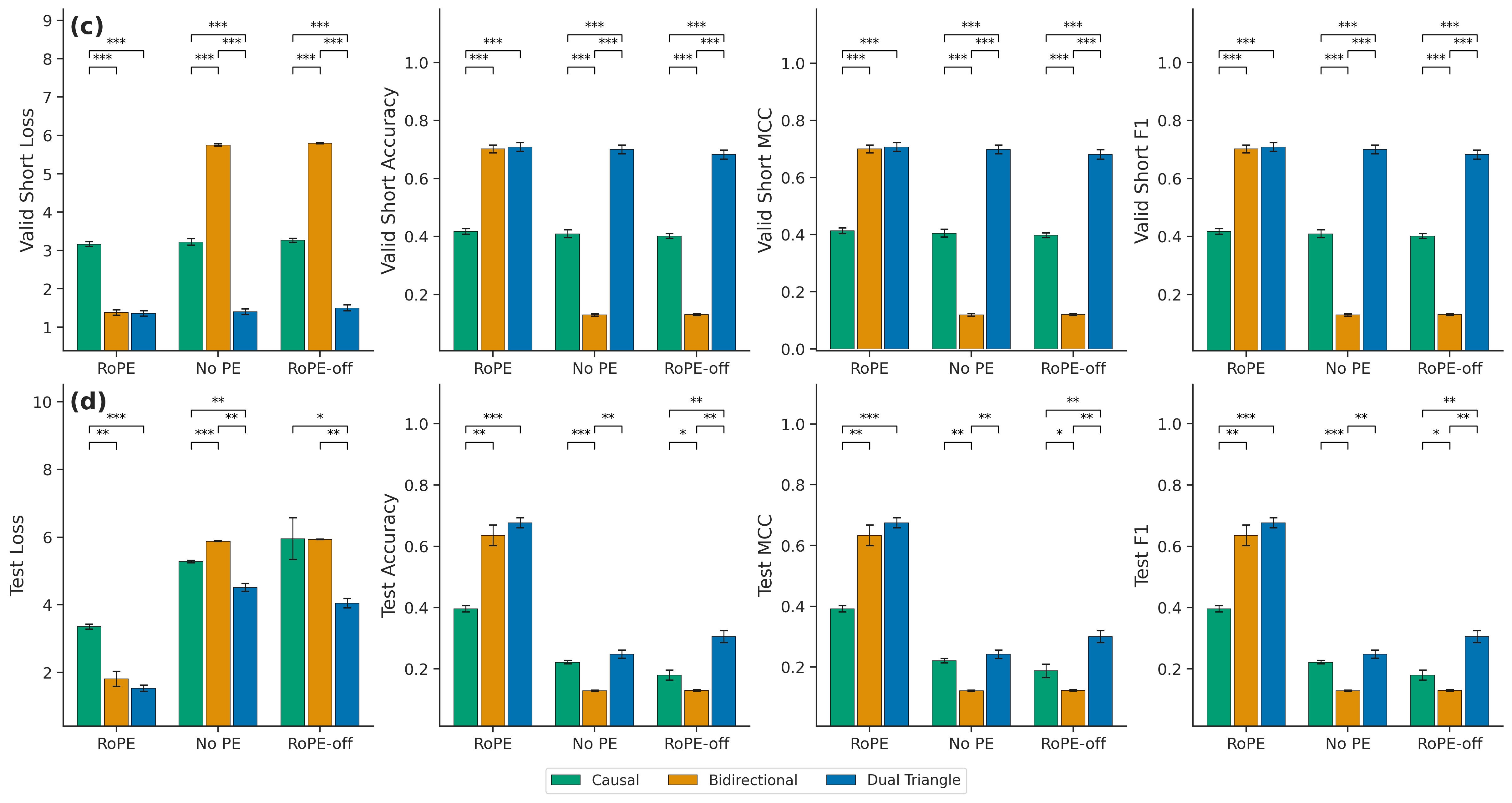}
  \caption{Masked language modeling on natural language (FineWeb-Edu). (a) Validation loss, accuracy, MCC, and F1 at training context length (256 tokens). (b) Validation loss, accuracy, MCC, and F1 at extended context length (1,024 tokens). Shaded regions represent $\pm$1 standard deviation across three seeds. (c) Best validation loss, accuracy, MCC, and F1 at training context length. (d) Test loss, accuracy, MCC, and F1 at extended context length. Significance brackets indicate Welch's $t$-test results: $^{*}$\,$p<0.05$, $^{**}$\,$p<0.01$, $^{***}$\,$p<0.001$. DTA + RoPE achieves the best performance at both training and extended context lengths. Without positional embeddings, DTA significantly outperforms bidirectional and causal attention.}
  \label{fig:nl_mlm}
\end{figure*}

At the training context length (\textbf{Figure \ref{fig:nl_mlm}a,c}), DTA with RoPE achieved the best validation loss (1.36 $\pm$ 0.07) and accuracy (0.709 $\pm$ 0.015), statistically comparable to bidirectional attention with RoPE (loss 1.38 $\pm$ 0.07, accuracy 0.702 $\pm$ 0.014; $p=0.68$, n.s.). Causal attention with RoPE performed substantially worse (loss 3.16 $\pm$ 0.06, accuracy 0.418 $\pm$ 0.010; $p < 0.001$), consistent with the expectation that unidirectional context is a significant limitation for MLM.

Without positional embeddings, DTA achieved a validation accuracy of 0.700 $\pm$ 0.015, remarkably close to its RoPE-enabled performance. In contrast, bidirectional attention without positional embeddings collapsed to 0.130 $\pm$ 0.004 accuracy ($p < 0.001$ vs.\ DTA; Cohen's $d > 50$), confirming that standard bidirectional attention cannot model language without explicit positional information. Causal attention without positional embeddings achieved 0.409 $\pm$ 0.013, demonstrating that the causal mask provides positional information, but unidirectional context limits MLM quality.

At the extended context length (4$\times$ training length), the performance hierarchy was even more pronounced (\textbf{Figure \ref{fig:nl_mlm}b,d}). DTA with RoPE achieved the best test loss (1.53 $\pm$ 0.09) and accuracy (0.677 $\pm$ 0.016), followed by bidirectional attention with RoPE (loss 1.81 $\pm$ 0.22, accuracy 0.636 $\pm$ 0.034; $p=0.15$, n.s.). Without positional embeddings, DTA maintained meaningful performance at extended context (test accuracy 0.248 $\pm$ 0.013), while bidirectional attention was essentially nonfunctional (0.129 $\pm$ 0.002; $p=0.004$; Cohen's $d = 12.8$).

The RoPE-off (DroPE) condition yielded mixed results for MLM. For DTA, dropping RoPE produced a test accuracy of 0.305 $\pm$ 0.019, which was significantly worse than the RoPE condition (0.677 $\pm$ 0.016; $p < 0.001$) but significantly better than the no-PE baseline at extended context ($p=0.004$). For bidirectional and causal attention, the RoPE-off condition severely degraded performance, with bidirectional attention collapsing to near the no-PE baseline (0.129 $\pm$ 0.002). These results suggest that while DroPE is effective for autoregressive models with causal masking \cite{gelberg_extending_2025}, it may not reliably transfer to the MLM setting, even for attention types with implicit positional structure.

\subsection*{Masked language modeling on protein sequences}

To assess the generality of our findings beyond natural language, we repeated the MLM experiments on protein sequences from the OMG-Prot50 dataset, using the ESM-2 tokenizer \cite{lin_evolutionary-scale_2023}. All architecture and training details were identical to those in natural language experiments, except for the tokenizer and data source.

\begin{figure*}[t]
  \centering
  \includegraphics[width=.78\textwidth]{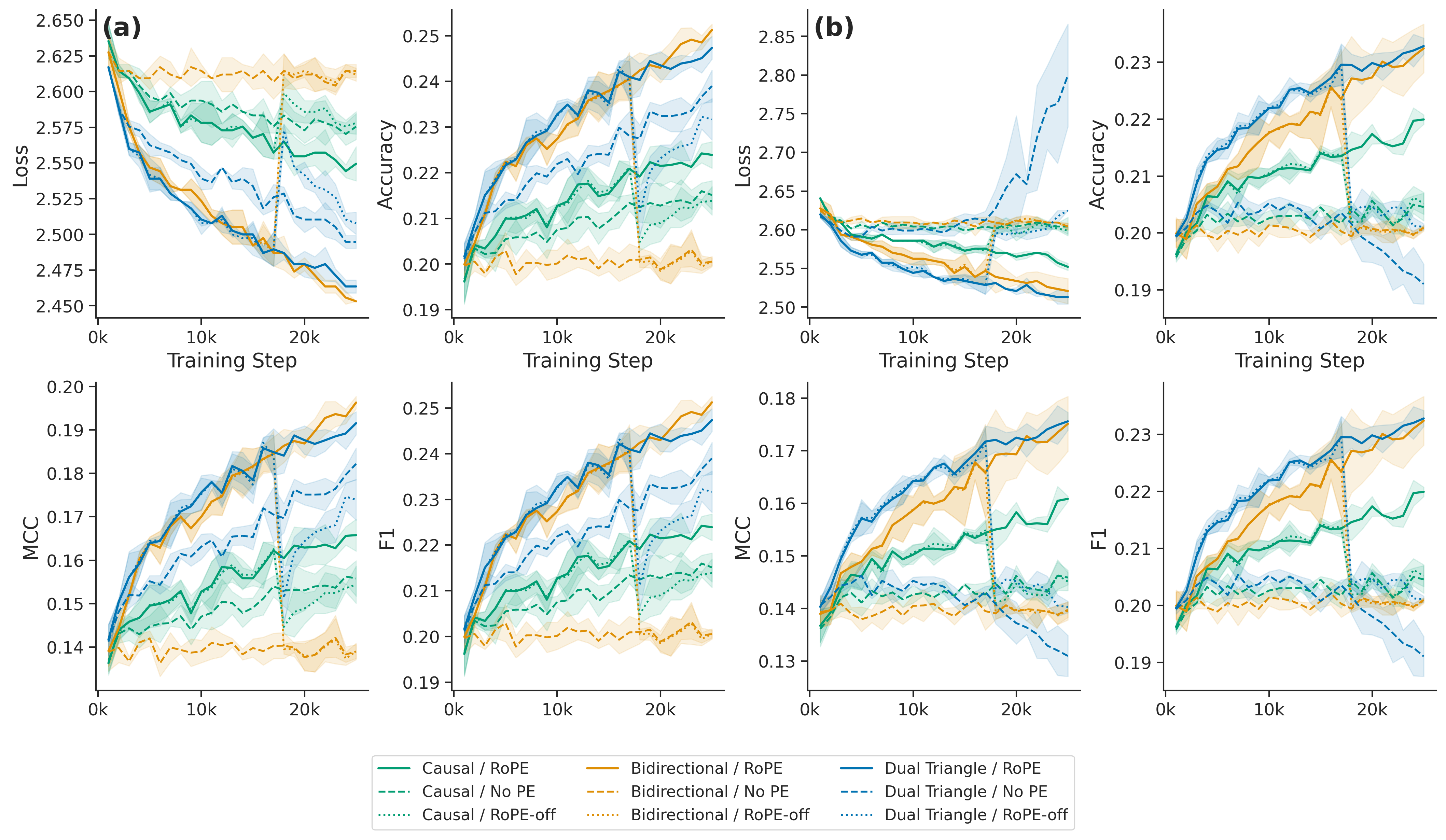}
  \vspace{0.5em}
  \includegraphics[width=.78\textwidth]{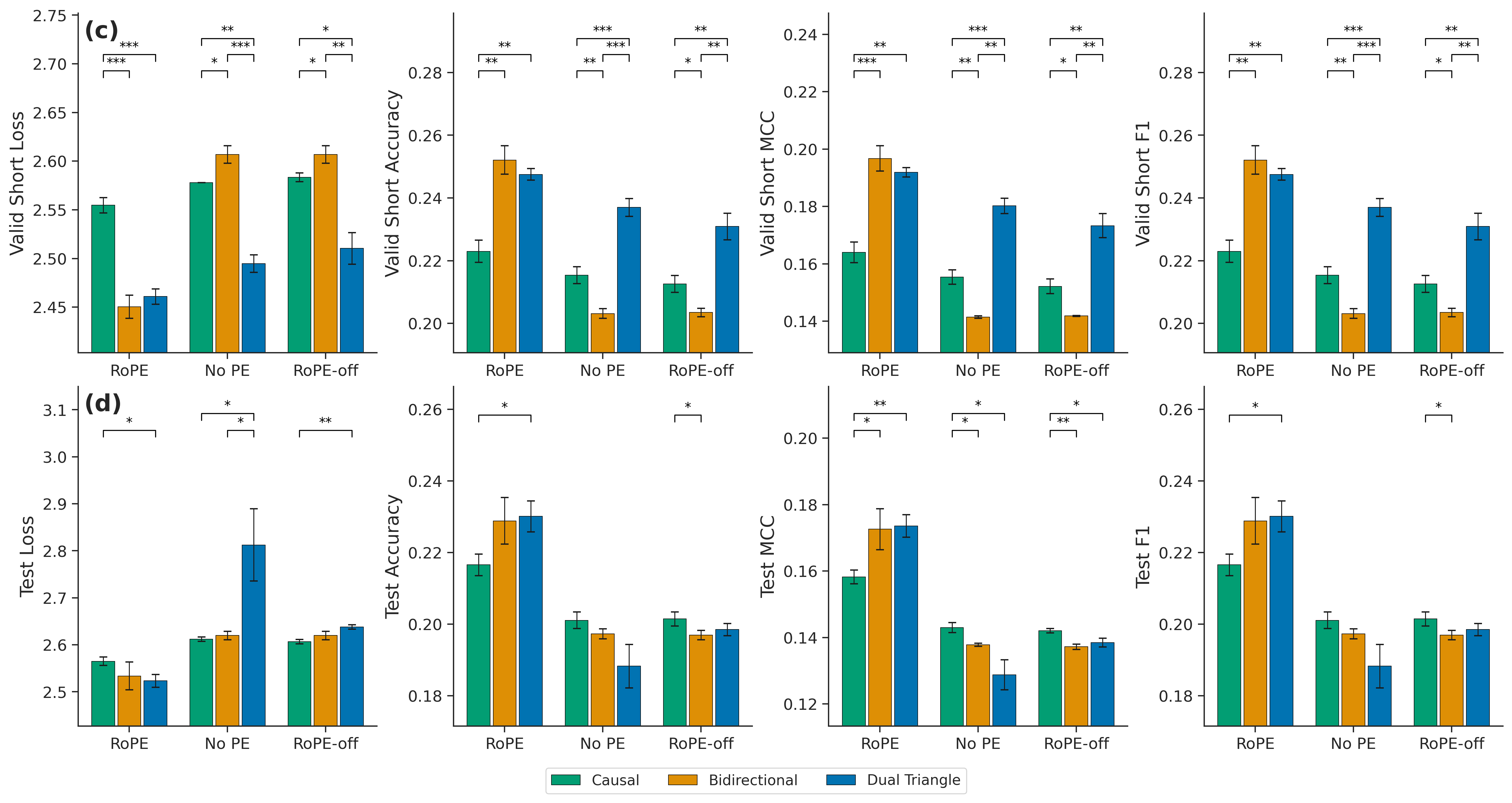}
  \caption{Masked language modeling on protein sequences (OMG-Prot50). (a) Validation loss, accuracy, MCC, and F1 at training context length (256 tokens). (b) Validation loss, accuracy, MCC, and F1 at extended context length (1,024 tokens). Shaded regions represent $\pm$1 standard deviation across three seeds. (c) Best validation loss, accuracy, MCC, and F1 at training context length. (d) Test loss, accuracy, MCC, and F1 at extended context length. Significance brackets indicate Welch's $t$-test results: $^{*}$\,$p<0.05$, $^{**}$\,$p<0.01$, $^{***}$\,$p<0.001$. Trends mirror the natural language experiments: DTA + RoPE performs best, and DTA without PE outperforms bidirectional without PE.}
  \label{fig:prot_mlm}
\end{figure*}

The overall trends mirrored the natural language experiments (\textbf{Figure \ref{fig:prot_mlm}}). At training context length, Dual Triangle Attention with RoPE achieved a competitive validation accuracy of 0.248 $\pm$ 0.002, comparable to bidirectional attention with RoPE at 0.252 $\pm$ 0.005 ($p=0.21$, n.s.) and significantly better than causal attention with RoPE (0.223 $\pm$ 0.004; $p=0.002$). At extended context, Dual Triangle Attention with RoPE achieved the best test loss (2.52 $\pm$ 0.01) and accuracy (0.230 $\pm$ 0.004), comparable to bidirectional with RoPE (loss 2.53 $\pm$ 0.03, accuracy 0.229 $\pm$ 0.007; $p=0.62$, n.s.) and significantly outperforming causal (loss 2.57 $\pm$ 0.01, accuracy 0.217 $\pm$ 0.003; $p=0.015$; Cohen's $d = 3.7$).

Without positional embeddings, Dual Triangle Attention again significantly outperformed bidirectional attention at training context length (validation accuracy 0.237 $\pm$ 0.003 vs.\ 0.203 $\pm$ 0.002; $p < 0.001$; Cohen's $d = 13.3$). Causal attention without positional embeddings performed competitively in this setting (0.215 $\pm$ 0.003).

\subsection*{DroPE does not reliably extend context for masked language modeling}

We examined whether the DroPE paradigm, training with RoPE and then removing it, could benefit MLM models, following the success of this approach in autoregressive language modeling \cite{gelberg_extending_2025}. In the RoPE-off condition, models were trained with RoPE for the first 70\% of the token budget (700M tokens) and then continued training without positional embeddings for the remaining 30\% (300M tokens).

\begin{figure*}[t]
  \centering
  \includegraphics[width=.78\textwidth]{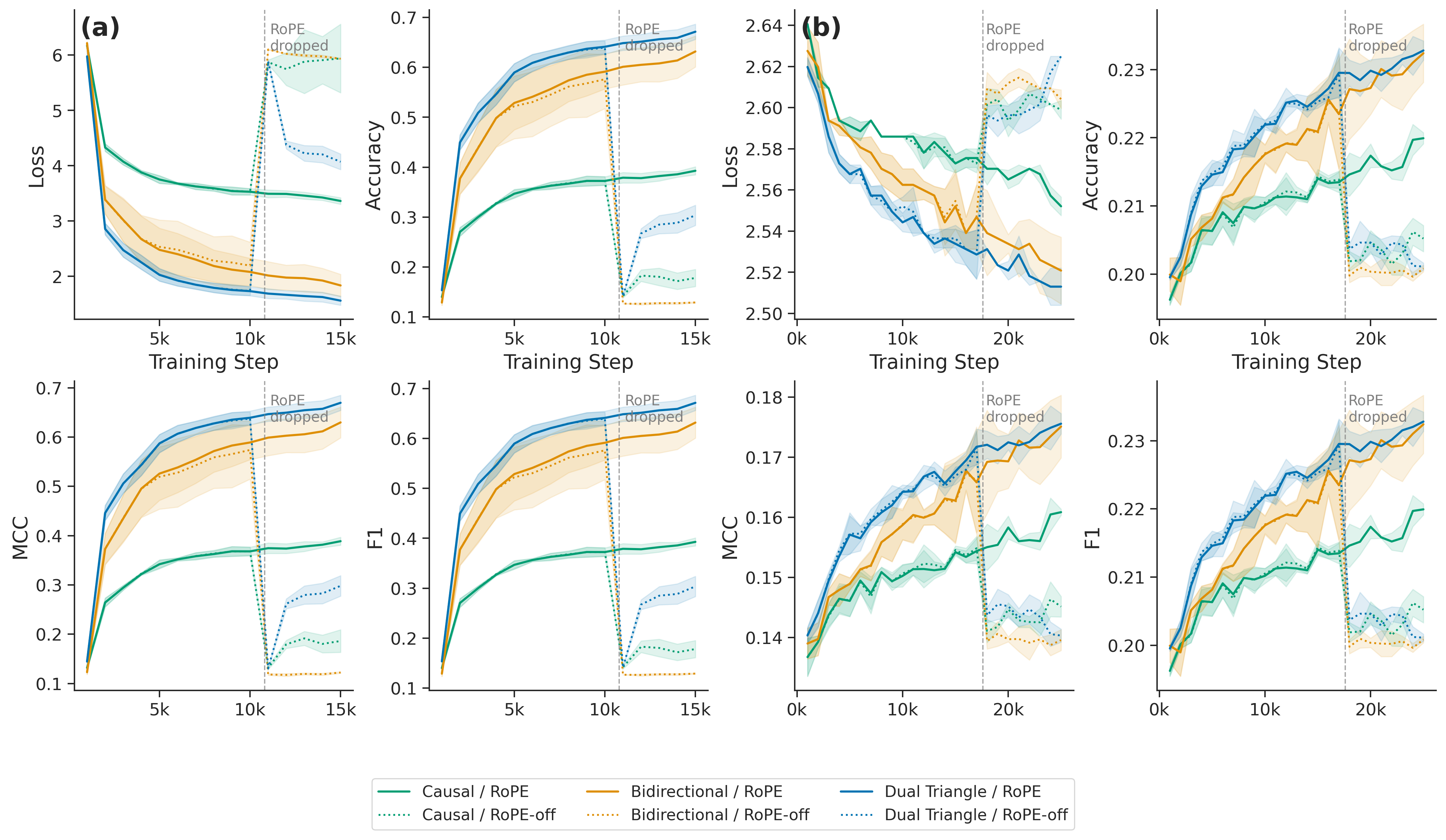}
  \vspace{0.5em}
  \includegraphics[width=.78\textwidth]{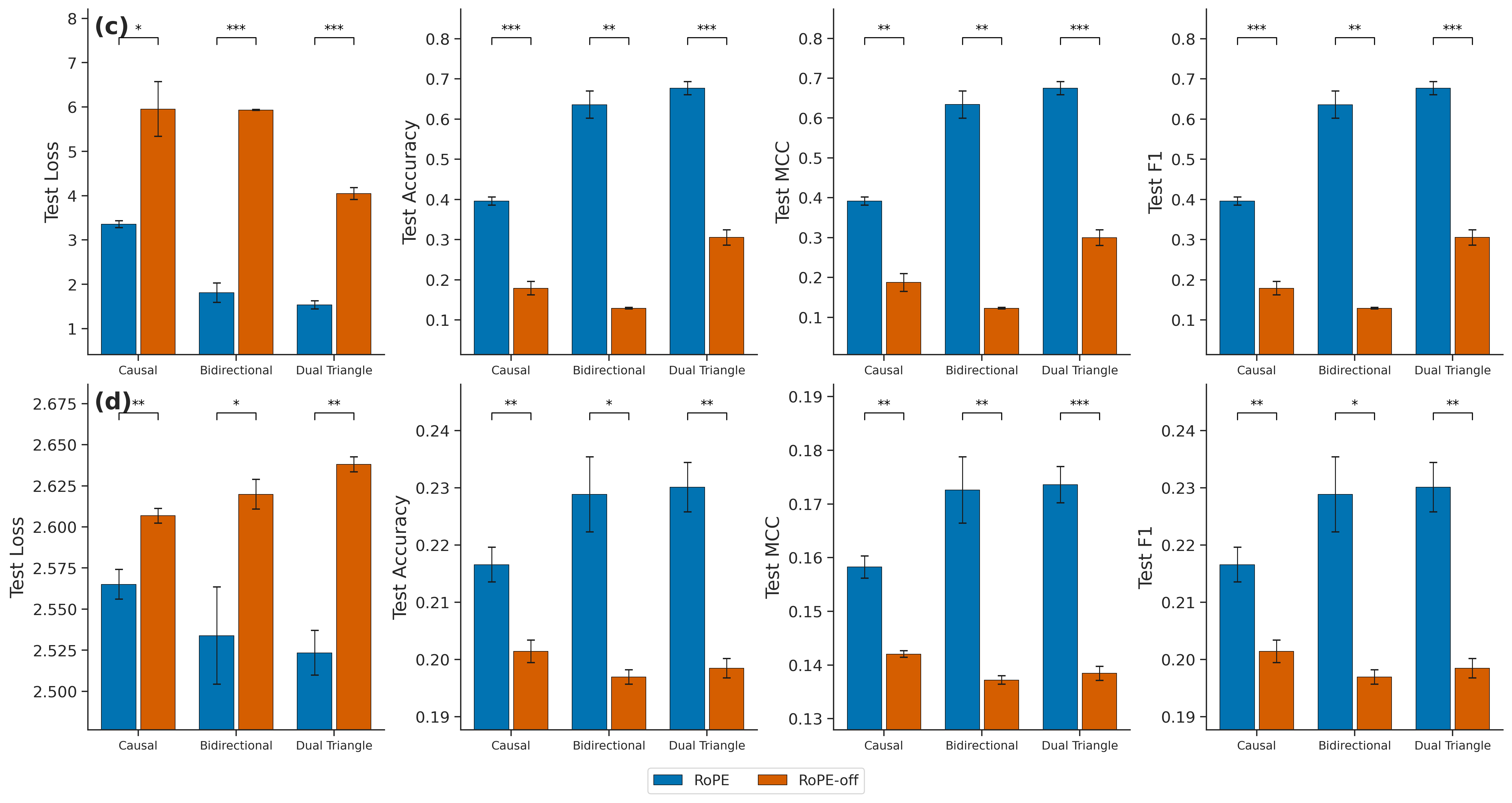}
  \caption{DroPE recovery analysis. (a) NLP extended-context validation loss, accuracy, MCC, and F1 before and after dropping positional embeddings at 70\% of training. (b) Protein extended-context validation loss, accuracy, MCC, and F1. The vertical dashed line marks the drop point. Shaded regions represent $\pm$1 standard deviation across three seeds. (c) NLP final test loss, accuracy, MCC, and F1 comparing RoPE (kept throughout) vs.\ RoPE-off (dropped at 70\%). (d) Protein final test loss, accuracy, MCC, and F1. Significance brackets indicate Welch's $t$-test results: $^{*}$\,$p<0.05$, $^{**}$\,$p<0.01$, $^{***}$\,$p<0.001$. Dropping positions degrades test performance across all attention types for both NLP and protein MLM, in contrast to the success of DroPE in autoregressive settings.}
  \label{fig:drop}
\end{figure*}

\textbf{Figure \ref{fig:drop}} shows the recovery dynamics and final metrics. Across both natural language and protein datasets, dropping positional embeddings led to a significant degradation in test performance relative to models that retained RoPE throughout training. On natural language, the test loss for Dual Triangle Attention increased from 1.53 $\pm$ 0.09 (RoPE) to 4.05 $\pm$ 0.14 (RoPE-off; $p=0.017$), and for bidirectional attention from 1.81 $\pm$ 0.22 to 5.93 $\pm$ 0.01 ($p=0.001$). Similar trends were observed on protein data, with Dual Triangle Attention test loss increasing from 2.52 $\pm$ 0.01 to 2.64 $\pm$ 0.005 ($p=0.006$) and causal from 2.57 $\pm$ 0.01 to 2.61 $\pm$ 0.005 ($p=0.002$).

Notably, Dual Triangle Attention exhibited the smallest degradation from position dropping in the NLP setting, with its training-context validation metrics (loss, accuracy, MCC) showing non-significant differences between RoPE and RoPE-off ($p > 0.10$). This partial resilience is consistent with the implicit positional signal provided by the triangular mask structure, but the extended-context degradation ($p=0.017$) indicates that this signal alone is insufficient to maintain performance at longer-than-training sequences after position dropping. We conclude that DroPE, in its current form, does not reliably transfer from the autoregressive to the MLM setting. This may be attributable to the bidirectional nature of MLM, where tokens must attend to both past and future context simultaneously, making the recalibration task fundamentally more complex than in the unidirectional case.

\section*{Discussion}

In this work, we established that Dual Triangle Attention (DTA) provides a general solution to the positional encoding problem in bidirectional attention. DTA encodes position without explicit PE, matches standard bidirectional attention with RoPE, achieves the best context extension performance across both NLP and protein domains, and reveals that DroPE does not transfer to the MLM setting. Our empirical findings are consistent with theoretical predictions about directional mask structures that allow models to implicitly learn position: Triangular masking, whether causal or dual-triangular, is sufficient for learning positional information without explicit positional embeddings, whereas symmetric bidirectional attention cannot learn position at all, regardless of model capacity. 

The argmax probe provides the cleanest demonstration of this principle. Because argmax prediction requires mapping token values to token positions, it serves as a direct test of positional awareness beyond perplexity-like metrics. The MLM results confirmed that this extends to practical tasks, as DTA without positional embeddings achieved 70\% validation accuracy on NLP, whereas bidirectional attention achieved only 13\%. DTA does not merely match bidirectional attention; it solves a problem that full bidirectional attention alone fundamentally cannot.

DTA with RoPE achieved the strongest context-extension performance across both domains, suggesting that implicit positional bias (from mask geometry) and explicit relative encoding (RoPE) are complementary rather than redundant. The triangular masks may provide coarse directional and ordinal structure, while RoPE supplies fine-grained relative distance information, consistent with head specialization patterns discussed by Barbero et al. \cite{barbero_round_2024}. This complementarity positions DTA within an emerging trend of hybrid PE/NoPE architectures. DTA represents the bidirectional instantiation of this hybrid approach, combining structural positional bias with explicit relative encoding in a single mechanism.

Our sweeps indicate that positional learning depends on model capacity: very small models (single transformer block, hidden size 4) failed on argmax, whereas large models (12 blocks, hidden size 768) were typically near-perfect, with intermediate variants showing a graded transition. One anomaly was the largest causal + RoPE configuration, which underperformed even though smaller causal + RoPE models succeeded. A plausible explanation is optimization instability from using a uniform learning rate ($3 \times 10^{-4}$) across all scales. This behavior was consistent across all three seeds for the large causal model and was not observed for the corresponding DTA variant.

Compellingly, DTA is competitive with, and in several settings superior to, standard bidirectional attention for sequence reconstruction. At training context length with RoPE, DTA matched bidirectional attention on both natural language (accuracy 0.709 vs.\ 0.702, $p=0.59$) and protein (0.248 vs.\ 0.252, $p=0.21$) MLM tasks. This parity is notable because DTA splits each head's query-key subspace between forward and backward directions. Although this split could reduce maximum per-direction expressiveness relative to vanilla multi-head attention, the observed performance suggests that directional specialization may compensate in practice. A further advantage is efficiency: the DTA attention operation requires approximately $\frac{1}{2}$ of the FLOPs of standard bidirectional attention because each direction uses half of the original query-key subspace.

Our results also indicate that DroPE does not transfer cleanly from autoregressive training to MLM. Although Gelberg et al. reported strong gains from position dropping in causal language models \cite{gelberg_extending_2025}, we observed substantial degradation when positions were dropped during MLM training, including for DTA. DroPE's success in causal models appears to rely on the structural information provided by the causal mask after positional embeddings are removed \cite{gelberg_extending_2025}. In bidirectional models, no comparable structural fallback is available, so removing positional embeddings from standard bidirectional attention severely disrupts training. The objective is also more demanding: in next-token prediction, the model only reconstructs forward context under an intact triangular mask, whereas MLM requires simultaneous integration of forward and backward evidence after positional removal. Practically, this suggests that context-extension strategies may need to differ between bidirectional and autoregressive architectures. The degradation was more pronounced for protein MLM, where no DroPE variant recovered to pre-drop language-modeling loss. In natural language, training loss often recovered during the final 30\% of training, but this recovery did not translate to validation or test performance; RoPE-off variants consistently trailed RoPE baselines on both short- and long-context evaluations.

We acknowledge several limitations. Most notably, our MLM architecture and optimizer choices were based on recent open-source engineering efforts that are not yet fully represented in the peer-reviewed literature. The U-Net-inspired transformer components, including explicit encoder/decoder paths, skip connections, and value-embedding injections, are non-standard. These design choices were motivated by recent progress from the Modded-NanoGPT and SpeedrunningPLMs communities \cite{jordan_kellerjordanmodded-nanogpt_2026, hallee_logan_gleghorn-labspeedrunningplms_2026}, including large reductions in training time and cost for transformer pretraining workflows. Likewise, the Muon optimizer has shown strong empirical performance in medium-to-large transformer regimes and has been used in some frontier-scale open-source training efforts \cite{noauthor_muon_nodate, team_kimi_2026}. Although these methods are relatively new, we applied them consistently across all attention variants to preserve fairness in the comparative conclusions.

Our MLM experiments used a fixed architecture (12 layers, 768 hidden size) and a single training budget (1B tokens), which is modest by modern standards. We chose these settings to enable thorough sweeps across attention types and positional conditions, but larger-scale experiments would be valuable to confirm whether the trends persist. The protein MLM results show smaller effect sizes than those for natural language, consistent with prior trends, suggesting that MLM on protein sequences is a harder task than MLM over natural language. Additionally, while we evaluated context extension up to 4$\times$ the training length, the training length was quite short at 256 tokens. Experiments with larger base training lengths and even larger extension factors would better characterize the scaling behavior of DTA and bidirectional DroPE.

Looking forward, DTA could serve as a drop-in replacement for standard bidirectional attention in masked language models and related sequence-understanding tasks. It has zero-parameter overhead, lower FLOPs, compatibility with compiled attention kernels, and built-in positional inductive bias, making it especially attractive for biological sequences, where long-range dependencies are central. In protein modeling, DTA combined with RoPE-scaling strategies such as YaRN \cite{peng_yarn_2023} may enable bidirectional models that train on typical protein lengths ($\approx$200-500 residues) but generalize to large multi-domain proteins and complexes (2,000-30,000+ residues) without expensive long-context retraining. For nucleotide modeling, where functional and structural dependencies can span large genomic distances, DTA's context extension capability may complement existing long-range approaches. Future work should focus on MLM-specific variants of position dropping, potentially enabling robust long-context extension in bidirectional settings without full long-context pretraining.


\section*{Methods}

\subsection*{Data sources}

\textbf{Argmax position probe.} Synthetic sequences were generated by sampling integers uniformly from $[0, v)$ where $v = 64$ is the vocabulary size. Labels were the 0-indexed position of the first occurrence of the maximum value. Sequence length was fixed at $l = 64$. Batches of 1,024 sequences were generated on-the-fly during training; evaluation used 16 batches of 1,024 sequences each.

\textbf{Natural language.} We used FineWeb-Edu \cite{penedo_fineweb_2024}, a large-scale filtered web corpus designed for language model pretraining. Text was tokenized using a custom Byte-Pair Encoding (BPE) tokenizer \cite{sennrich_neural_2016} with a vocabulary of 4,096 tokens, chosen to reduce vocabulary size relative to standard tokenizers while preserving reasonable subword granularity. Training sequences were truncated or padded to 256 tokens. Validation and test sets were constructed by filtering documents with at least 1,024 tokens, then splitting the remaining documents into 1,000 documents each for validation and testing. Training data was streamed and filtered to exclude validation and test documents.

\textbf{Protein sequences.} We used the OMG-Prot50 dataset, which contains protein sequences clustered at 50\% sequence identity. Sequences were tokenized using the ESM-2 tokenizer \cite{lin_evolutionary-scale_2023}, which uses a standard amino acid vocabulary. All other data pipeline details were identical to those in the natural language setting.

\subsection*{Attention mechanisms}

We evaluated three attention types, all implemented using PyTorch's \texttt{flex\_attention} API \cite{dong_flex_2024} with compiled block masks for hardware-efficient execution.

\textbf{Bidirectional attention.} Standard full attention with no masking. Every token attends to every other token:
\begin{equation}
\text{mask}(q_i, k_j) = \text{True} \quad \forall \, i, j
\end{equation}

\textbf{Causal attention.} Standard autoregressive masking where each token attends only to itself and preceding tokens:
\begin{equation}
\text{mask}(q_i, k_j) = (i \geq j)
\end{equation}

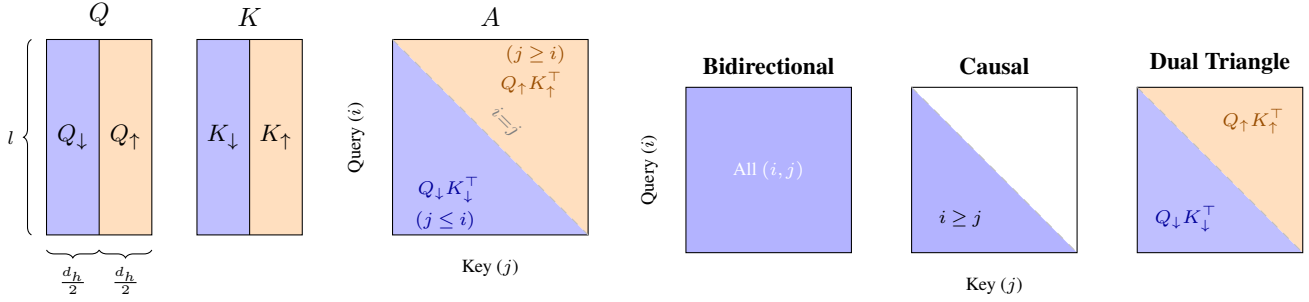
\begin{figure*}[ht]
  \centering
  \begin{subfigure}[t]{0.42\textwidth}
    \centering
    \begin{tikzpicture}[>=Stealth]
      \fill[blue!25] (0, 0) rectangle (0.7, 2.6);
      \fill[orange!25] (0.7, 0) rectangle (1.4, 2.6);
      \draw (0, 0) rectangle (1.4, 2.6);
      \draw (0.7, 0) -- (0.7, 2.6);
      \node[font=\small] at (0.35, 1.3) {$Q_\downarrow$};
      \node[font=\small] at (1.05, 1.3) {$Q_\uparrow$};
      \node[above, font=\normalsize\bfseries] at (0.7, 2.65) {$Q$};
      \fill[blue!25] (2.0, 0) rectangle (2.7, 2.6);
      \fill[orange!25] (2.7, 0) rectangle (3.4, 2.6);
      \draw (2.0, 0) rectangle (3.4, 2.6);
      \draw (2.7, 0) -- (2.7, 2.6);
      \node[font=\small] at (2.35, 1.3) {$K_\downarrow$};
      \node[font=\small] at (3.05, 1.3) {$K_\uparrow$};
      \node[above, font=\normalsize\bfseries] at (2.7, 2.65) {$K$};
      \draw[decorate, decoration={brace, amplitude=3pt, mirror}]
        (0, -0.2) -- (0.7, -0.2)
        node[midway, below, yshift=-3pt, font=\scriptsize] {$\frac{d_h}{2}$};
      \draw[decorate, decoration={brace, amplitude=3pt, mirror}]
        (0.7, -0.2) -- (1.4, -0.2)
        node[midway, below, yshift=-3pt, font=\scriptsize] {$\frac{d_h}{2}$};
      \draw[decorate, decoration={brace, amplitude=3pt}]
        (-0.2, 0) -- (-0.2, 2.6)
        node[midway, left, xshift=-3pt, font=\scriptsize] {$l$};
      \pgfmathsetmacro{\mx}{4.6}
      \pgfmathsetmacro{\ms}{2.6}
      \fill[blue!25] (\mx, 0) -- (\mx+\ms, 0) -- (\mx, \ms) -- cycle;
      \fill[orange!25] (\mx, \ms) -- (\mx+\ms, \ms) -- (\mx+\ms, 0) -- cycle;
      \draw (\mx, 0) rectangle (\mx+\ms, \ms);
      \draw[gray!50, dashed] (\mx, \ms) -- (\mx+\ms, 0);
      \node[font=\scriptsize, blue!60!black] at (\mx+0.75, 0.6) {$Q_\downarrow K_\downarrow^\top$};
      \node[font=\scriptsize, blue!60!black] at (\mx+0.70, 0.2) {$(j \leq i)$};
      \node[font=\scriptsize, orange!60!black] at (\mx+\ms-0.75, \ms-0.6) {$Q_\uparrow K_\uparrow^\top$};
      \node[font=\scriptsize, orange!60!black] at (\mx+\ms-0.70, \ms-0.2) {$(j \geq i)$};
      \node[font=\scriptsize, gray, rotate=-45] at (\mx+\ms/2+0.2, \ms/2+0.2) {$i{=}j$};
      \node[above, font=\normalsize\bfseries] at (\mx+\ms/2, \ms+0.05) {$A$};
      \node[below, font=\scriptsize] at (\mx+\ms/2, -0.2) {Key ($j$)};
      \node[rotate=90, anchor=south, font=\scriptsize] at (\mx-0.25, \ms/2) {Query ($i$)};
    \end{tikzpicture}
    \caption{Query--key tensor decomposition for one logical head.}
    \label{fig:dt_qk}
  \end{subfigure}
  \hfill
  \begin{subfigure}[t]{0.55\textwidth}
    \centering
    \begin{tikzpicture}[>=Stealth]
      \pgfmathsetmacro{\ms}{2.2}
      \pgfmathsetmacro{\bx}{0}
      \fill[blue!30] (\bx, 0) rectangle (\bx+\ms, \ms);
      \draw (\bx, 0) rectangle (\bx+\ms, \ms);
      \node[above, font=\small\bfseries] at (\bx+\ms/2, \ms+0.05) {Bidirectional};
      \node[font=\scriptsize, white] at (\bx+\ms/2, \ms/2) {All $(i,j)$};
      \pgfmathsetmacro{\cx}{3.0}
      \fill[blue!30] (\cx, 0) -- (\cx+\ms, 0) -- (\cx, \ms) -- cycle;
      \draw (\cx, 0) rectangle (\cx+\ms, \ms);
      \draw[gray!50, dashed] (\cx, \ms) -- (\cx+\ms, 0);
      \node[above, font=\small\bfseries] at (\cx+\ms/2, \ms+0.05) {Causal};
      \node[font=\scriptsize] at (\cx+0.65, 0.45) {$i \geq j$};
      \pgfmathsetmacro{\dx}{6.0}
      \fill[blue!25] (\dx, 0) -- (\dx+\ms, 0) -- (\dx, \ms) -- cycle;
      \fill[orange!25] (\dx, \ms) -- (\dx+\ms, \ms) -- (\dx+\ms, 0) -- cycle;
      \draw (\dx, 0) rectangle (\dx+\ms, \ms);
      \draw[gray!50, dashed] (\dx, \ms) -- (\dx+\ms, 0);
      \node[above, font=\small\bfseries] at (\dx+\ms/2, \ms+0.05) {Dual Triangle};
      \node[font=\scriptsize, blue!60!black] at (\dx+0.65, 0.45) {$Q_\downarrow K_\downarrow^\top$};
      \node[font=\scriptsize, orange!60!black] at (\dx+\ms-0.65, \ms-0.45) {$Q_\uparrow K_\uparrow^\top$};
      \node[below, font=\scriptsize] at (\cx+\ms/2, -0.2) {Key ($j$)};
      \node[rotate=90, anchor=south, font=\scriptsize] at (\bx-0.25, \ms/2) {Query ($i$)};
    \end{tikzpicture}
    \caption{Attention masks with $Q/K$ subspace annotations.}
    \label{fig:dt_patterns}
  \end{subfigure}
  \caption{Dual Triangle Attention. (\textbf{a}) Each logical head's $Q$ and $K$ tensors ($l \times d_h$) are split at $d_h/2$: the down half ($Q_\downarrow, K_\downarrow$, blue) computes the lower triangle ($j \leq i$) and the up half ($Q_\uparrow, K_\uparrow$, orange) computes the upper triangle ($j \geq i$). The diagonal ($i = j$) is shared by both sub-heads. (\textbf{b}) Attention mask comparison. Bidirectional attention permits all position pairs. Causal attention masks future tokens. Dual Triangle Attention provides full bidirectional coverage through complementary triangular masks from separate $Q$/$K$ subspaces, via a single \texttt{flex\_attention} call with head-routed block masks.}
  \label{fig:dt_attention_detail}
\end{figure*}

\textbf{Dual Triangle Attention.} Each logical attention head with dimension $d_{\text{head}}$ is split into two sub-heads, each with dimension $d_{\text{head}} / 2$. The first sub-head (``down'') attends to the lower triangle (past and self), and the second sub-head (``up'') attends to the upper triangle (future and self):
\begin{equation}
\text{mask}(h, q_i, k_j) =
\begin{cases}
i \geq j & \text{if } h < n \text{ (down sub-head)} \\
i \leq j & \text{if } h \geq n \text{ (up sub-head)}
\end{cases}
\end{equation}
where $n$ is the number of logical heads. The diagonal ($i = j$) is included in both triangles to ensure no attention row is empty, which would cause numerical issues in the softmax normalization. Each sub-head receives independent softmax normalization. The query--key tensor decomposition and resulting attention patterns are illustrated in \textbf{Figure \ref{fig:dt_attention_detail}}.

In practice, the full implementation reshapes $Q, K, V \in \mathbb{R}^{b \times n \times l \times d_{\text{head}}}$ to $\mathbb{R}^{b \times 2n \times l \times d_{\text{head}}/2}$ by splitting each head's dimension in half, applies the head-routed block mask via a single \texttt{flex\_attention} call, and reshapes the output back. If used, RoPE was applied to the full $d_{\text{head}}$-dimensional $Q$ and $K$ vectors \textit{before} the split, ensuring that rotational phases are consistent across the two sub-heads. Pseudocode for both the naive and \texttt{flex\_attention} implementations is provided in the Supplemental Material.

\subsection*{Positional embedding strategies}

\textbf{None.} No positional information is provided; the model relies solely on the attention mask structure.

\textbf{Learned absolute.} A learnable embedding matrix $P \in \mathbb{R}^{l \times d}$ is added to token embeddings before the first transformer block, following the standard BERT convention \cite{devlin_bert_2019}.

\textbf{RoPE.} Rotary positional embeddings \cite{su_roformer_2023} are applied to query and key vectors at each attention layer. We use the standard parameterization with base frequency $b = 10{,}000$:
\begin{equation}
\omega_m = b^{-2(m-1)/d_k}, \quad m = 1, \ldots, d_k / 2
\end{equation}
where $d_k$ is the head dimension.

\textbf{DroPE-style (RoPE-off / learned-abs-off).} Models are trained with RoPE (or learned absolute) for the first 70\% of the token budget, after which positional embeddings were disabled, and training continued without them. This was inspired by DroPE \cite{gelberg_extending_2025}, adapted to the MLM setting.

\subsection*{Transformer architectures}

\textbf{Argmax probe.} We used a standard pre-norm transformer encoder. Positional embeddings (if used) were added to the token embeddings. The output was processed through a pooling head: per-position logits were computed, passed through softmax to produce position weights, and used to compute a weighted sum of hidden states, which was then classified over position indices via a linear layer. Hidden sizes of 4, 64, and 768 were evaluated with 1, 4, and 12 layers, respectively. Head size was 64 for bidirectional and causal attention, and 128 for Dual Triangle (to ensure the $d_{\text{head}}/2 = 64$ sub-heads have sufficient capacity). Dropout was set to 0.

\begin{figure*}[ht]
  \centering
  \begin{tikzpicture}[
    >=Stealth,
    block/.style={rectangle, draw, rounded corners=3pt, minimum width=2.8cm,
                  minimum height=0.8cm, align=center, font=\small},
    enc/.style={block, fill=blue!8},
    dec/.style={block, fill=orange!8},
    io/.style={block, fill=gray!10},
    flow/.style={->, thick},
    skip/.style={->, dashed, thick, gray!60!black},
    ve/.style={->, thin, teal!60!black},
    x0arr/.style={->, dotted, thin, violet!40!black},
  ]
    \node[io] (emb) at (0, 7) {Token Emb + Pos Emb$_{(optional)}$};
    \node[io] (head) at (7.6, 7) {LM Head};
    \node[enc] (e1) at (0, 4.8) {Encoder Block 1};
    \node[enc] (e2) at (0, 2.5) {Encoder Block 2};
    \node[dec] (d3) at (7.6, 2.5) {Decoder Block 3};
    \node[dec] (d4) at (7.6, 4.8) {Decoder Block 4};
    \draw[flow] (emb) -- (e1);
    \draw[flow] (e1) -- (e2);
    \draw[flow] (e2.south) -- ++(0, -0.7) -| (d3.south);
    \draw[flow] (d3) -- (d4);
    \draw[flow] (d4) -- (head);
    \draw[skip] (e1.east) -- node[above, font=\scriptsize] {$w_1 \cdot \text{skip}_1$} (d4.west);
    \draw[skip] (e2.east) -- node[above, font=\scriptsize] {$w_2 \cdot \text{skip}_2$} (d3.west);
    \node[font=\scriptsize, teal!60!black] (ve1l) at (-2.6, 4.8) {$\text{ve}_1(x)$};
    \node[font=\scriptsize, teal!60!black] (ve2l) at (-2.6, 2.5) {$\text{ve}_2(x)$};
    \node[font=\scriptsize, teal!60!black] (ve2r) at (10.2, 2.5) {$\text{ve}_{n-1}(x)$};
    \node[font=\scriptsize, teal!60!black] (ve1r) at (10.2, 4.8) {$\text{ve}_n(x)$};
    \draw[ve] (ve1l) -- (e1.west);
    \draw[ve] (ve2l) -- (e2.west);
    \draw[ve] (ve2r) -- (d3.east);
    \draw[ve] (ve1r) -- (d4.east);
    \node[font=\scriptsize, violet!50!black, fill=white, draw=violet!30,
          rounded corners=2pt, inner sep=2pt] (x0) at (3.8, 6.0) {$x_0$};
    \draw[->, thin, violet!40!black] (0.2, 5.9) -- (x0.west);
    \node[font=\tiny, violet!50!black, anchor=north] at (3.8, 5.55)
      {$\lambda_0 h + \lambda_1 x_0$ at each block};
    \draw[x0arr] (x0.south west) to[out=210, in=80] ([xshift=-5pt]e1.north);
    \draw[x0arr] (x0.south west) to[out=225, in=55] ([xshift=-5pt]e2.north east);
    \draw[x0arr] (x0.south east) to[out=-30, in=100] ([xshift=5pt]d4.north);
    \draw[x0arr] (x0.south east) to[out=-45, in=125] ([xshift=5pt]d3.north west);
    \node[font=\scriptsize, blue!50!black, rotate=90, anchor=south] at (-1.8, 3.65) {Encoder};
    \node[font=\scriptsize, orange!50!black, rotate=-90, anchor=south] at (9.4, 3.65) {Decoder};
  \end{tikzpicture}
  \caption{U-Net transformer architecture for masked language modeling. The encoder ($\frac{n_{\text{layers}}}{2}$ blocks) feeds forward through the left path, and the decoder ($\frac{n_{\text{layers}}}{2}$ blocks) ascends through the right path. Dashed arrows indicate skip connections with learnable scalar weights $w_{\text{skip}}$. Per-layer value embeddings $\text{ve}(x)$ (\texttt{nn.Embedding}) from raw token indices are injected into each block. The original input representation $x_0$ is mixed into every block via learnable scalars $\lambda_0, \lambda_1$. Shown with four blocks; generalizes to any even $n_{\text{layers}}$.}
  \label{fig:U-Net_arch}
\end{figure*}
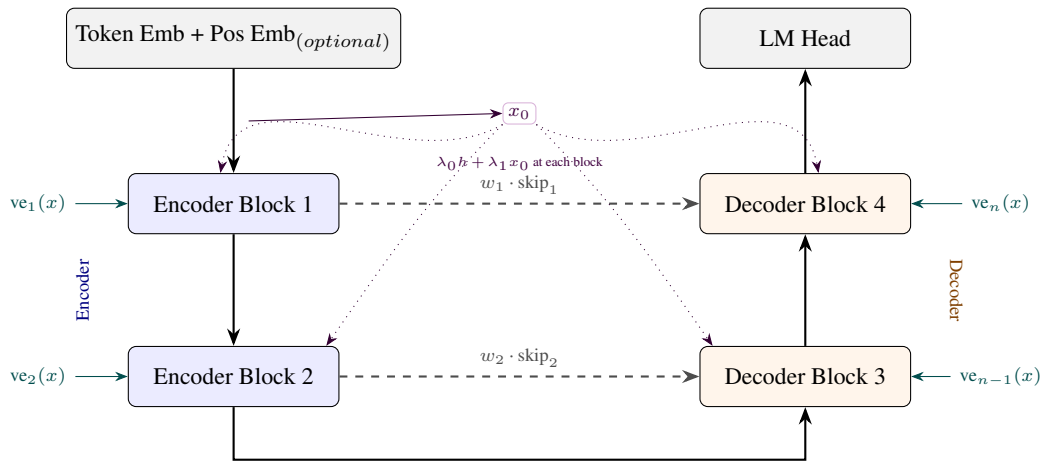

\textbf{MLM (U-Net transformer).} For the masked language modeling experiments, we used a U-Net-style transformer architecture inspired by recent work on efficient pretraining \cite{bao_all_2023, jordan_kellerjordanmodded-nanogpt_2026}. The architecture consists of $n_{\text{layers}}/2$ encoder layers followed by $n_{\text{layers}}/2$ decoder layers, with skip connections from each encoder layer to the corresponding decoder layer:
\begin{equation}
h_{\text{dec}}^{(r)} = \text{Block}(h_{\text{dec}}^{(r-1)}) + w_{\text{skip}}^{(r)} \cdot h_{\text{enc}}^{(n_{\text{layers}}/2 - r)}
\end{equation}
where $w_{\text{skip}}^{(r)}$ is a learnable scalar weight initialized to 1. Additionally, each block mixes the original input $x_0$ (the sum of token and positional embeddings) into the residual stream via learnable lambdas:
\begin{equation}
\tilde{h}^{(r)} = \lambda_0^{(r)} \cdot h^{(r)} + \lambda_1^{(r)} \cdot x_0
\end{equation}
Value embeddings, consisting of per-layer learnable embeddings from raw input token indices, were also mixed into the transformer blocks. The language modeling head consisted of a two-layer MLP with GELU activation \cite{hendrycks_gaussian_2023} followed by a linear projection to the vocabulary. All models used $n_{\text{layers}} = 12$ layers, hidden size 768, SwiGLU MLPs \cite{shazeer_glu_2020}, pre-norm LayerNorm \cite{ba_layer_2016}, and no dropout. Head size was 64 for bidirectional and causal models, and 128 for the Dual Triangle model. The complete U-Net architecture is illustrated in \textbf{Figure \ref{fig:U-Net_arch}}.

\subsection*{Training details}

\textbf{Argmax probe.} Models were trained with AdamW \cite{loshchilov_decoupled_2019} (learning rate $3 \times 10^{-4}$, weight decay 0.01) using a linear warmup followed by cosine decay schedule. Training proceeded in evaluation cycles of 256 steps each, with early stopping after 3 consecutive evaluations without improvement, up to a maximum of 10 evaluation cycles. Models were cast to bfloat16 for forward and backward passes.

\textbf{MLM.} Models were trained with a joint Muon + AdamW optimizer \cite{noauthor_muon_nodate, loshchilov_decoupled_2019, team_kimi_2026}. Muon (learning rate 0.01) was applied to 2D weight matrices (attention projections, MLP weights) using Newton-Schulz orthogonalization of momentum updates. AdamW (learning rate $1 \times 10^{-3}$, weight decay 0) was applied to embeddings, layer norms, biases, and scalar parameters. The learning rate schedule consisted of a linear warmup (100M tokens), constant phase (800M tokens), and cosine cooldown (100M tokens), for a total of 1B non-padding tokens. The batch size was 256. Models were cast to bfloat16 \cite{micikevicius_mixed_2018} without autocast or gradient scaling, following the approach of recent speedrunning efforts \cite{jordan_kellerjordanmodded-nanogpt_2026}. MLM masking used the standard 80/10/10 scheme (80\% mask token, 10\% random token, 10\% unchanged) with a masking probability of 15\%, applied on the GPU to minimize CPU-GPU synchronization. Three seeds (11, 22, 33) were used for each condition.

\subsection*{Evaluation and statistical analysis}

Argmax models were evaluated on prediction accuracy (proportion of sequences where the predicted position matched the true argmax position). MLM models were evaluated on cross-entropy loss, token-level accuracy, micro-averaged F1 score, and Matthews Correlation Coefficient (MCC) \cite{matthews_comparison_1975}, computed over masked positions only. Short-context evaluation used the training sequence length (256 tokens) and long-context evaluation used 4$\times$ the training length (1,024 tokens). Final test metrics were computed on a held-out set of 1,000 documents. 

Statistical comparisons between attention types used two-sided Welch's $t$-tests \cite{welch_generalization_1947} across seeds ($n = 3$). We report $p$-values with standard significance thresholds: $^*$ $p < 0.05$, $^{**}$ $p < 0.01$, $^{***}$ $p < 0.001$, n.s.\ $p \geq 0.05$. DroPE recovery was assessed by comparing the same attention type between RoPE and RoPE-off conditions.


\section*{Data and code availability}

All code, training configurations, and analysis scripts are available at \url{https://github.com/Gleghorn-Lab/DualTriangleAttention}. Training data sources are publicly available via Huggingface: FineWeb-Edu \cite{penedo_fineweb_2024} (\url{https://huggingface.co/datasets/HuggingFaceFW/fineweb-edu}) | OMG-Prot50 \cite{cornman_omg_2024} Original (\url{https://huggingface.co/datasets/tattabio/OMG_prot50}) with Train Valid Test splits (\url{https://huggingface.co/datasets/Synthyra/omg_prot50}).


\section*{Acknowledgements}

The authors thank Katherine M.\ Nelson, Ph.D. and Nikolaos Rafailidis for reviewing and commenting on drafts of the manuscript.

\section*{Funding}

This work was partly supported by the University of Delaware Graduate College through the Unidel Distinguished Graduate Scholar Award (LH), the National Science Foundation through NAIRR pilot 240064 (JPG), and the National Institutes of Health through NIGMS T32GM142603 (LH), R01HL178817 (JPG), and R01HL133163 (JPG).

\section*{Conflict of interest}

LH and JPG are co-founders of and have an equity stake in Synthyra, PBLLC.

\section*{Author contributions}

Conceived and designed the experiments (LH), performed the experiments (LH), analyzed the data (LH), contributed materials/analysis tools (LH), wrote the paper (LH, JPG), supervised the work (JPG), funding acquisition (LH, JPG).


\section*{References}
\bibliography{references}

\newpage


\section*{Supplementary information}

\subsection*{Dual Triangle Attention pseudocode}

We provide pseudocode for both a naive implementation and the hardware-efficient \texttt{flex\_attention} implementation.

\begin{algorithm}[H]
\caption{Dual Triangle Attention (Naive)}
\label{alg:naive}
\begin{algorithmic}[1]
\Require Input $X \in \mathbb{R}^{b \times l \times d}$, weight matrices $W_Q, W_K, W_V, W_O$
\State $Q, K, V \gets XW_Q,\; XW_K,\; XW_V$ \Comment{Project to $\mathbb{R}^{b \times l \times d}$}
\State Reshape $Q, K, V$ to $\mathbb{R}^{b \times n \times l \times d_h}$ \Comment{$n$ heads, head dim $d_h$}
\If{RoPE enabled}
    \State $Q, K \gets \text{apply\_rope}(Q, K)$
\EndIf
\State $Q_\downarrow, Q_\uparrow \gets Q[\ldots, :d_h/2],\; Q[\ldots, d_h/2:]$ \Comment{Split heads}
\State $K_\downarrow, K_\uparrow \gets K[\ldots, :d_h/2],\; K[\ldots, d_h/2:]$
\State $s = (d_h / 2)^{-1/2}$
\State $A_\downarrow \gets Q_\downarrow K_\downarrow^\top \cdot s$ \Comment{Past + self logits}
\State $A_\uparrow \gets Q_\uparrow K_\uparrow^\top \cdot s$ \Comment{Future + self logits}
\State $M_\text{lower} \gets \text{tril}(\mathbf{1}_{l \times l})$ \Comment{Lower triangle (incl.\ diagonal)}
\State $A \gets M_\text{lower} \odot A_\downarrow + (1 - M_\text{lower} + \text{diag}(\mathbf{1}_l)) \odot A_\uparrow$
\State Mask padding positions with $-\infty$ if applicable
\State $\alpha \gets \text{softmax}(A, \text{dim}=-1)$
\State $O \gets \alpha V$ \Comment{$\mathbb{R}^{b \times n \times l \times d_h}$}
\State Reshape $O$ to $\mathbb{R}^{b \times l \times d}$, project: $\text{output} \gets OW_O$
\end{algorithmic}
\end{algorithm}

\begin{algorithm}[H]
\caption{Dual Triangle Attention (Flex Attention)}
\label{alg:flex}
\begin{algorithmic}[1]
\Require Input $X \in \mathbb{R}^{b \times l \times d}$, weight matrices $W_Q, W_K, W_V, W_O$
\State $Q, K, V \gets XW_Q,\; XW_K,\; XW_V$
\State Reshape to $\mathbb{R}^{b \times n \times l \times d_h}$ \Comment{$n$ logical heads}
\If{RoPE enabled}
    \State $Q, K \gets \text{apply\_rope}(Q, K)$ \Comment{Full $d_h$ rotation}
\EndIf
\State Reshape to $\mathbb{R}^{b \times 2n \times l \times d_h/2}$ \Comment{Split each head into 2 sub-heads}
\State \textbf{Define} \textsc{mask\_mod}$(b, h, q_i, k_j)$:
\State \quad $\texttt{is\_down} \gets h < n$
\State \quad \Return $(\texttt{is\_down} \wedge q_i \geq k_j) \vee (\neg\texttt{is\_down} \wedge q_i \leq k_j)$
\State $M \gets \texttt{create\_block\_mask}(\textsc{mask\_mod}, b, 2n, l, l)$
\State $O \gets \texttt{flex\_attention}(Q, K, V, \text{block\_mask}=M)$ \Comment{Single kernel call}
\State Reshape $O$ back to $\mathbb{R}^{b \times n \times l \times d_h}$, then to $\mathbb{R}^{b \times l \times d}$
\State $\text{output} \gets OW_O$
\end{algorithmic}
\end{algorithm}

The \texttt{flex\_attention} implementation (Algorithm \ref{alg:flex}) is preferred because: (1) it avoids materializing the full $l \times l$ attention matrix, (2) the block mask is compiled into an efficient Triton kernel, and (3) only one kernel call is needed regardless of the number of sub-heads. The naive implementation (Algorithm \ref{alg:naive}) is provided for clarity and for frameworks that do not support \texttt{flex\_attention}.

\subsection*{Argmax experiment controls}

\begin{figure}[H]
  \centering
  \includegraphics[width=0.5\textwidth]{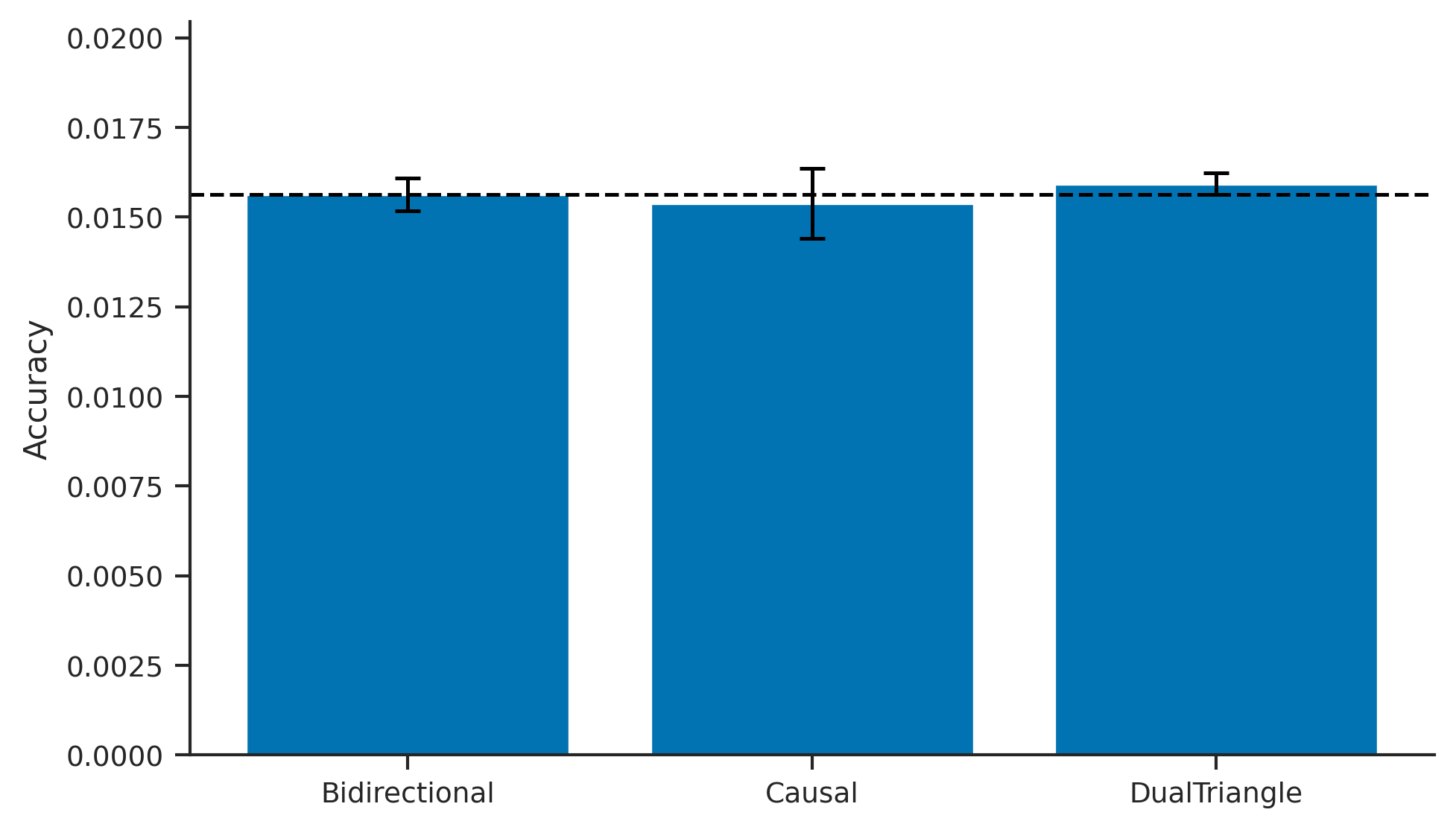}
  \caption{Argmax control experiment with random labels. All attention types achieve near-chance accuracy when labels are randomized, confirming that high accuracy in the main experiments reflects genuine positional learning.}
  \label{sup_fig:control}
\end{figure}

To confirm that the argmax models were learning positional information rather than exploiting spurious correlations, we trained control models with \textbf{random labels} (uniformly sampled positions, independent of the input sequence). All model configurations achieved near-chance accuracy ($\sim 1/l = 1.6\%$) on random labels (\textbf{Supplemental Figure \ref{sup_fig:control}}), confirming that the main experiment results reflect genuine learning of the position-value association.

\end{document}